\documentclass[aps,prl,superscriptaddress,showpacs]{revtex4}
\newcommand{\be}{\begin{equation}}
\newcommand{\ee}{\end{equation}}
\newcommand{\ba}{\begin{eqnarray}}
\newcommand{\ea}{\end{eqnarray}}

\newcommand{\wAM}{\omega_{\scriptscriptstyle\rm AM}}

\usepackage{color}
\usepackage{graphicx}
\usepackage{bm}
\usepackage{epsfig}
\usepackage{subfigure}
\usepackage{subfigure}
\usepackage{graphicx}
\begin{document}


\title{Heterogeneous Dynamics, Marginal Stability and Soft Modes in Hard Sphere Glasses}

\author{Carolina Brito}
\affiliation{Instituto de F\'{\i}sica, Universidade Federal do
 Rio Grande do Sul, Av. Bento Golçalves 9500, 91501-970 Porto Alegre, Brazil}
\author {Matthieu Wyart}
\affiliation{Division of Engineering and Applied Sciences,
Harvard University, Pierce Hall,
29 Oxford Street, Cambridge,
 Massachusetts 02138, USA}


\date{\today}

\begin{abstract}

In a recent publication we established an analogy between the free energy of a
hard sphere system and the energy of an elastic network \cite{caca}. This 
 result  enables one to  
study the free energy landscape of hard spheres, in particular to define normal modes. 
In this Letter we use these tools to analyze the activated
transitions between meta-bassins,  both in the aging regime deep
 in the glass phase and near the glass transition.  We observe numerically that structural 
relaxation occurs mostly along a very small number  of nearly-unstable
extended modes. This number decays for denser packing and is significantly lowered as the system undergoes the glass transition. This observation supports that structural relaxation and marginal modes share common properties. In particular theoretical results
\cite{matthieu1,matthieu2} show that  these modes  extend at least on
some length scale $l^*\sim (\phi_c-\phi)^{-1/2}$ where $\phi_c$
corresponds to the maximum packing fraction, i.e. the jamming
transition. This prediction  is consistent with  very recent numerical
observations of sheared systems near the  jamming threshold \cite{cond}, where a similar exponent is found, and with the commonly  observed growth of the rearranging  regions with compression near the glass transition. 

\end{abstract}

\maketitle

A colossal effort has been made to characterize the spatial nature of the
structural relaxation  near the glass transition.
 Numerical simulations \cite{simu} and experiments \cite{exp, weitz} have
 shown that the dynamics in super-cooled liquids is heterogeneous. Both the
 string-like \cite{glotzer} and the compact \cite{kob}  aspects of the
 particles' displacements have 
 been emphasized. Nevertheless, the cause
 of such collective motions remains debated \cite{toni, tarjus}. To make progress,
 one would like to relate these motions to other objects.  A
 possible candidate is the excess of low-frequency modes present in all
 glasses, the so-called  boson peak \cite{phillips_book}. Because these modes
 shift in general to lower frequencies as the temperature 
 increases toward  the 
 glass transition temperature $T_g$, it has been proposed that they are
 responsible for the melting of the glass \cite{alexander,parisi}. 
This  suggests  the use of widely employed  tools, 
 such as the low-frequency  instantaneous normal modes \cite{keyes} 
or the negative directions of saddles  of the potential energy
landscape  \cite{saddle}, to analyze the collective
 motions causing relaxation.
Nevertheless, this approach has the major
 drawback of being  based on energy instead of free energy. 
 As such, it cannot be applied for example to hard spheres or colloids, where structural
 relaxation is also known to be collective, see e.g. \cite{weitz}. In this case barriers
 between meta-stable states are purely entropic. More generally, one expects
 entropic effects to be important for glasses where hard-core repulsions and
 non-linearities are not negligible, which is presumably  the case in general above 
 $T_g$ \cite{chandler}.

Recent developments make this analysis possible in hard sphere systems. In
 \cite{caca}, we derived an analogy between the free energy of a hard sphere
 glass  and the energy of a weakly-connected
 network of logarithmic springs. This allows us to define normal modes, that can
 be compared  with the dynamics.  Furthermore,  recent results
 \cite{matthieu1,matthieu2,matthieu3} valid for weakly-connected networks, such as
 elastic particles near jamming \cite{J,sil} ---where scaling laws between packing geometry and vibrational properties were first observed--- or  simple models of silica 
 \cite{dove},
apply to characterize these  modes:  (i) Excess modes appear above some
 frequency $\wAM$  which depends on the pressure $p$ and the
 coordination $z$, whose definition shall be recalled below for hard
 spheres. These {\it anomalous modes} 
 extend at least on a length scale $l^*$, which depends on $z$ and diverges 
 near maximum packing \cite{matthieu2,matthieu3,sil}.  (ii) Meta-stable states
 can exist only if they contain a configuration for which
 $\wAM > 0$.  This leads to a non-trivial
 scaling  relationship  between $p$  and  $z$ that must be
 satisfied in the glass phase.   Numerically, we observed that the hard sphere
 glass lies close to marginal stability: the coordination is just sufficient
 to  maintain rigidity \cite{caca}. This implies that anomalous modes
 are present at very low-frequency.

In this Letter, we study how low-frequency modes
 take part in the structural relaxation, both during the aging dynamics deep
 in the glass phase, and in the vicinity of the glass transition where the
 system is  at equilibrium.
 We show that when relaxation occurs between meta-stable states, the
 system yields in the direction of the softest  modes: most of the
 amplitude of the observed  displacements can be decomposed on a  small
 fraction of the modes, of the order of few percent. This observation supports
 that the collective aspect of the relaxation
 does not stem from  the non-linear coupling of localized relaxation 
 events, 
  but rather from the extended character of the softest degrees of freedom.
  This suggests that the typical size of the events relaxing the structure increases as 
  the extension of the anomalous modes $l^*\sim (\phi_c-\phi)^{-1/2}\sim p^{1/2}$, which diverges deep in the glass phase.

We start by recalling some results of \cite{caca}. 
In a meta-stable state of a 
hard sphere system, one  can define  a contact network \cite{bubble}: 
two particles are said to be in contact if they
collide during some interval of time $t_1$, where $t_1$ is chosen to be much
larger than an $\tau_c$ , the collision time,  and  smaller than the structural
relaxation time $\tau$ where meta-stability is lost. 
The coordination number $z$ of this network is defined as 
the average number of contacts of the particles in the system.
An approximation of the Gibbs free energy ${\cal G}$ can then be expressed 
as a sum on all the contacts $\langle ij\rangle$: 
 \be
 \label{1}
 {\cal G}= - kT \sum_{\langle ij\rangle} \ln (\langle h_{ij}  \rangle_t)
 \ee
 where $h_{ij}= r_{ij} -r_i-r_j$ is the gap between particles $i$ and $j$,
 $r_{ij}$ is the distance between  them,  $r_i$ denotes the radius of particle
 $i$, and $\langle\rangle_t$ is a time-average.  Eq.(\ref{1}) has two main
 limitations:  (i) it is only exact near the maximum packing fraction
 $\phi_c$  where the pressure
 diverges and (ii) to perform the time-average one requires a strong
 separation of time scales between $\tau_c$ and $\tau$. 
Thus Eq.(\ref{1}) is  a better approximation deep in the glass phase.
 Nevertheless the corrections  to Eq.(\ref{1}) are found to be rather small
 empirically\cite{caca,matthieu3},  and we shall use Eq.(\ref{1}) to 
study the vicinity of the glass transition ($\phi\approx \phi_0$) as well.

 Eq.(\ref{1}) can be expanded around any equilibrium
 position. For a contact $ij$, one finds for the force
 $V'_{ij}=-kT/\langle h_{ij}\rangle_t$ and for the stiffness
 $V''_{ij}=kT/\langle h_{ij}\rangle_t^2$. 
This enables one to compute the dynamical matrix ${\cal M}$ \cite{as}
 which relates a small applied force to the linear displacement of the average
 particle positions. Normal modes can then be computed, whose angular
 frequencies are the square  roots  of the eigenvalues of ${\cal M}$.  
In what follows we locate quiet periods of the dynamics
 where  ${\cal M}$ can be estimated. Then, we use the normal modes to
 analyze the subsequent structural  relaxation.

\begin{figure}                
  \begin{center}
    \rotatebox{0}{\resizebox{5.5cm}{!}{\includegraphics{f1a.eps}}}
    \rotatebox{0}{\resizebox{6.9cm}{!}{\includegraphics{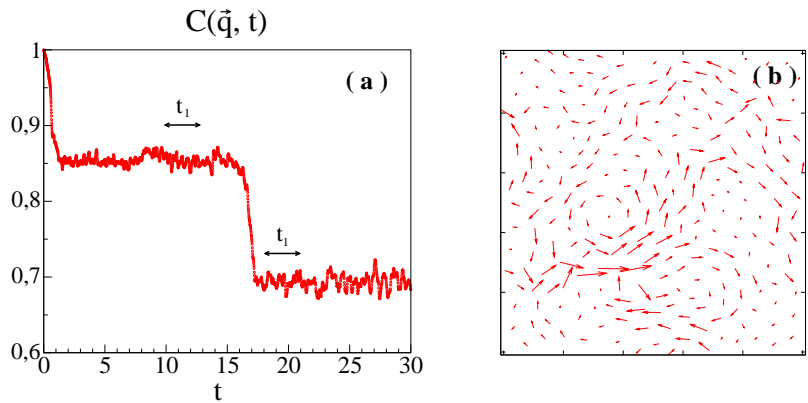}}}
    \caption{
      Left: Self-density correlation function $C({\vec q}, t)$ {\it vs.} time for 
      $q= 2\pi/\sigma_1$ in a system of $N=256$ particles, at packing fraction
      $\phi=0.837$. Meta-stable states appear as plateaus of $C(\vec q,t)$, whereas the
      drops of $C(\vec q,t)$ are the mentioned earthquakes. 
      Time-averages are made during the time segments $t_1$.  Right: Displacement
      field of the corresponding earthquake. Arrows connect the average  
      particles position  before and after the earthquake, they are 
      amplified 4 times here for visibility.   For similar data in a 3d LJ see \cite{kb}.
    }
    \label{ex_plat_crack}
\end{center}
\end{figure}

 We consider a  bidisperse  two-dimensional hard sphere
 system. Half of the particles have a diameter $\sigma_1=1$, the other a
 diameter 1.4, their mass is $m=1$, and energies are expressed in units of
 $kT$. To study the aging dynamics,  configurations are generated in the glass
 phase ($\phi_0\approx 0.79 \leq \phi\leq \phi_c \approx 0.84$)  as in
 \cite{caca}. An event-driven code is used to simulate the dynamics. 
We  observe long quiet
 periods, or meta-stable states, interrupted
 by sudden rearrangements, or ``earthquakes".  Such earthquakes correspond to
 collective motions of a large number of particles, and  have been
 observed in various other aging systems, such as colloidal paste or laponite \cite{duri}, and in
 Lennard-Jones simulations \cite{lapo,kb}.  Even for our largest numerical box of $N=1024$ particles, deep in the glass phase these events generally  span the entire system. They appear as drops in the self
 scattering function $C(\vec q, t)\equiv \langle exp [i {\vec q} . ({\vec R_i}(t) -
{\vec R_i}(0) )]\rangle_i$, where $\langle \rangle_i$ is an average on all
 particles  and  $\vec R_i(t)$ is the  position of particle $i$ at time $t$. 
An example of earthquake is shown in  Fig.(\ref{ex_plat_crack}). 

 In what follows the average particles position in a meta-stable state $l$ is 
denoted $ |R^l\rangle \equiv  \{ \langle{\vec R_i}\rangle_t \}, i=1...N$.
In practice the time-average $\langle \rangle_t$  is done on a long 
time $t_1$  corresponding to few hundred collisions per particle 
(we use  $t_1=10^5, \ 5\times10^4$  
numerical time steps for  respectively $N=1024$ and $ 256$ particles).
 The earthquake displacement field $|\delta R^e\rangle$ between two
 meta-stable states $l$ and $m$ is then defined as 
$|\delta R^e\rangle\equiv |R^m\rangle- |R^l\rangle$, see
Fig(\ref{ex_plat_crack}-b).
During earthquakes, we find that the average particles 
displacement is typically 10\%-20\% of the particles diameter,
 and tend to decrease with the pressure.

\begin{figure}   
\vspace{1.3cm}
  \begin{center}
    \rotatebox{0}{\resizebox{5.5cm}{!}{\includegraphics{f2a.eps}}}
    \rotatebox{0}{\resizebox{6.9cm}{!}{\includegraphics{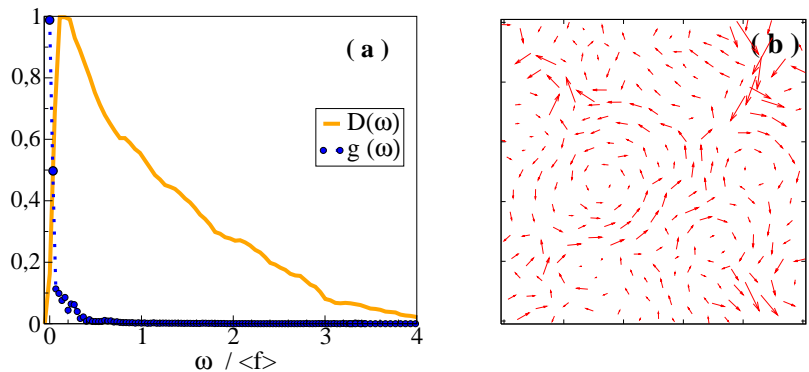}}}              
   \caption{Left, straight curve:  $D(\omega)$  computed in
     the meta-stable state prior to the earthquake of Fig.(\ref{ex_plat_crack})
     {\it vs.}  $\omega/ \langle f\rangle$, the angular frequency rescaled 
     by the average contact force $\langle f\rangle$, computed by averaging
     on all pairs of particles in contact.  Dotted curve: 
     $g(\omega)$ {\it vs.} $\omega/\langle f\rangle$. Right:
     Lowest frequency mode.}
   \label{Dw_proj}
 \end{center}
\end{figure}

To analyze these displacement fields, we compute the average  of the  
 particles' positions and the contact network  in the meta-stable state prior to the
earthquake\footnote[1]{Very close to $\phi_c$ (for $\langle f\rangle>5 \times
 10^3$), 
 ``rattlers''   \cite{J} are present, which are systematically removed from our analysis \cite{caca}. }.
This enables us to define ${\cal M}$
 and the normal modes $|\delta R^\alpha\rangle$, where the label
 $\alpha=1,...,2N$  ranks the modes by
increasing frequencies  $\omega^{\alpha}$ . An example of the  density of states $D(\omega)$
 is shown in Fig.(\ref{Dw_proj}-a), together with the lowest frequency mode, 
Fig.(\ref{Dw_proj}-b).
We indeed observe extended anomalous modes at very low  frequencies, in agreement
 with the marginal  stability inferred from the microscopic structure of the glass \cite{caca}.
Note that we occasionally observe a few unstable modes even deep in the glass
 phase,  implying the presence of saddles (and multiple configurations of free 
 energy minima) or ``shoulders" in the meta-stable states we are considering. 
In the present work we do not focus on this aspect, and treat unstable modes as the rest.

We then project the earthquake  displacement $|\delta R^e\rangle$ 
on the  modes and compute
 $c_\alpha=\langle \delta R^e|\delta R^\alpha\rangle/ \langle \delta R^e|
 \delta R^e\rangle$, where 
$\langle \delta R^e |\delta  R^\alpha\rangle\equiv \sum_i \delta\vec{ R_i^e}\cdot
 \delta\vec{R_i^\alpha}$. 
The $c_\alpha$'s  satisfy $\sum_\alpha c_\alpha{}^2=1$ since the normal modes form a unitary basis. 
 To study how the contribution of the modes  depends on 
 frequency, we define $g(\omega)= \langle c_\alpha{}^2\rangle_\omega$, where
 the average is made on all $\alpha$ such that $\omega^\alpha\in [\omega,
 \omega+d\omega]$. Fig.(\ref{Dw_proj}-a)  shows $g(\omega)$ for the earthquake
 shown in Fig.(\ref{ex_plat_crack}). The average contribution of
 the modes decreases very rapidly with increasing frequency, and  most of the displacement projects on the excess-modes present near zero-frequency. This implies that 
 the free energy barrier crossed by the system is located in the direction of the softest degrees of freedom.

\begin{figure}                                
  \begin{center} 
    \rotatebox{-90}{\resizebox{8.0cm}{!}{\includegraphics{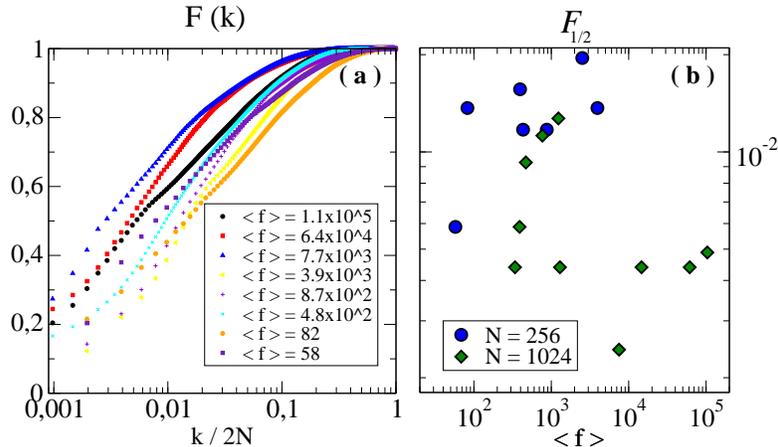}}}
    \caption{Left: Examples of $F(k)$ {\it vs.} $k/2N$ for systems with different average contact force
      $\langle f \rangle$.  $\langle f \rangle$ is measured before the
      earthquake.  It is  proportional to the
      pressure near $\phi_c$, and is of the order of  20 near the  glass transition $\phi_0$.   
      Right:  $F_{1/2}$ {\it vs.}    $\langle f \rangle$ for  $N=256$ (circles) and
      $N=1024$ (diamonds) particles.}
    \label{Fk_alpha2}
  \end{center}
\end{figure}

To make this observation systematic, we introduce the label $i$ to rank the
 $c$'s  by decreasing order: $c_1>c_2...>c_{2N}$. Then we define:
\be
F(k)\equiv \sum_{i=1}^k c_i^2
\ee
$F(k)$  indicates  which fraction of the total displacement is contained in
the $k$ most  contributing modes. If $F(k)=1\ \forall k$ then only one mode contributes.
If $F(k)=k/2N$ all modes contribute equally. We then define $k_{1/2}$ as the
minimum number of modes contributing to $50\%$ of the displacements, i.e.  the
smallest $k$ for which $F(k)>1/2$. Fig.(\ref{Fk_alpha2})  shows
 $F(k)$ and $F_{1/2}\equiv k_{1/2}/(2N)$ for the 17 cracks 
studied.  Fig.(\ref{Fk_alpha2}-b) shows that  $0.2\%<F_{1/2}< 2\%$  for all the events studied throughout  the glass phase.
We thus systematically observe that the extended earthquakes correspond  to the relaxation of a  small number  of degrees of freedom.

\begin{figure}
  \begin{center}
    \hspace{-0.5cm}
    \rotatebox{-90}{\resizebox{4.4cm}{!}{\includegraphics{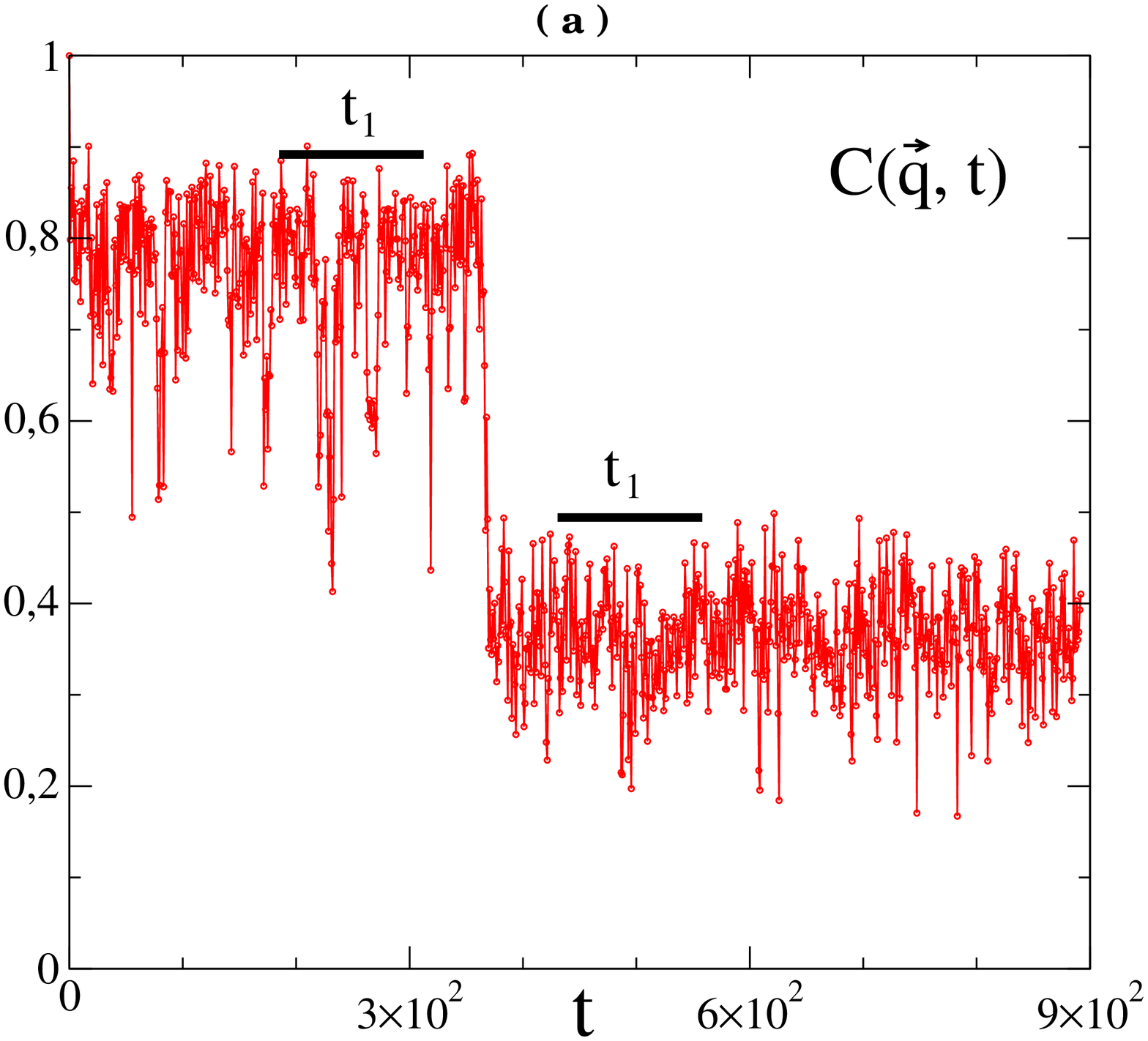}}}
    \hspace{-0.5cm}
    \rotatebox{-90}{\resizebox{4.4cm}{!}{\includegraphics{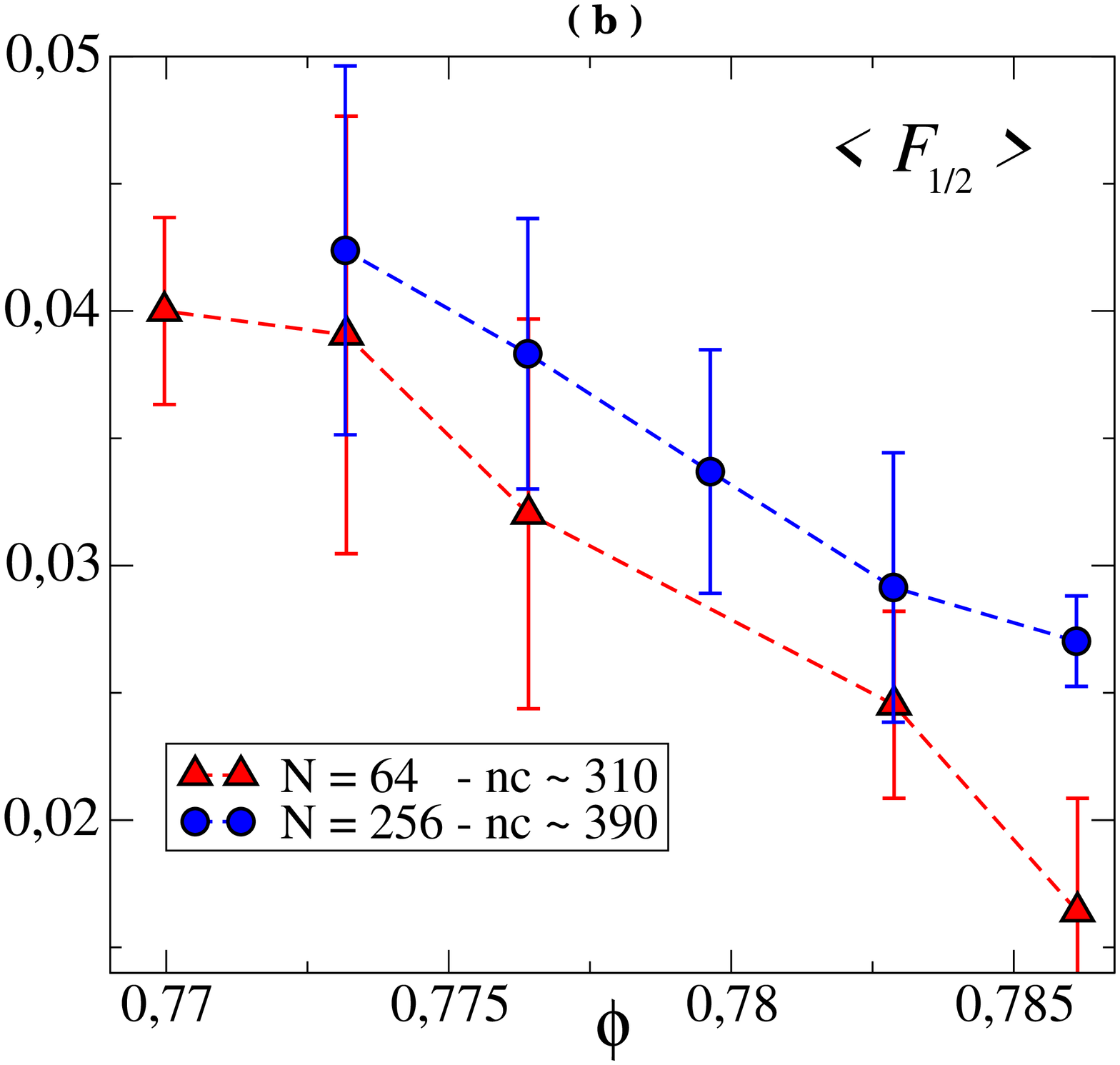}}}
    \hspace{-0.5cm}
    \rotatebox{-90}{\resizebox{4.4cm}{!}{\includegraphics{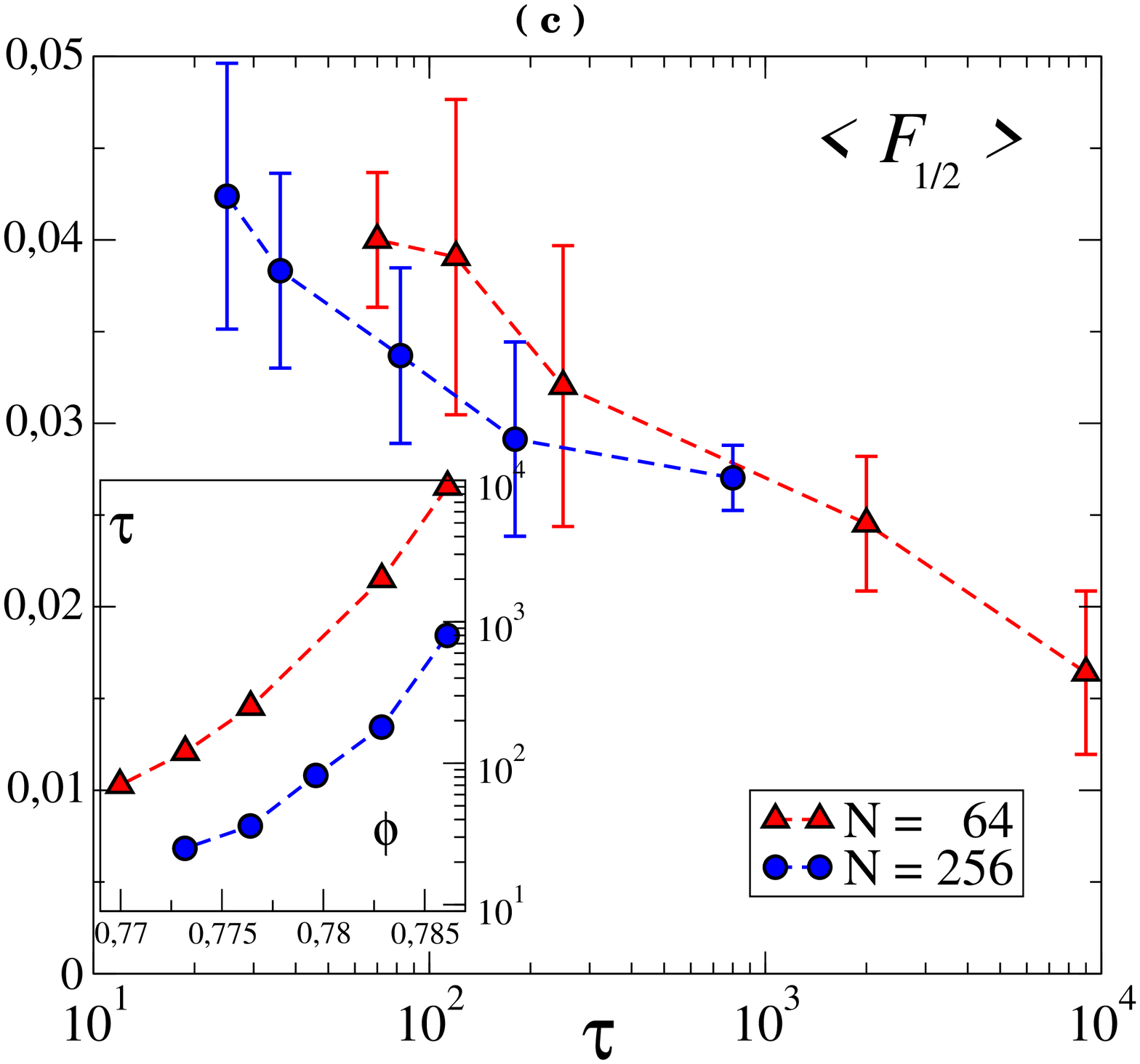}}}
    \caption{{\bf (a)}: $C(\vec q, t)$ in an equilibrated system of $N=64$
      particles at $\phi=0.786$. The segment $t_1$ corresponds to
      $10^4$ numerical  time steps,  which corresponds roughly to 300
      collisions   on average per particle.
      {\bf(b)}: $\langle F_{1/2} \rangle$  {\it vs}  $\phi$ for 
      $t_1=5\times 10^4$ (circle) for $N=256$  and 
       $t_1=10^4$ (triangle)  for a system with $N=64$. The average number of
      collisions per particle is written in the legend and denoted by $nc$.
      {\bf(c)}:
      $\langle F_{1/2} \rangle$  {\it vs} $\tau$ for two different
      system size.  Inset: relaxation time $\tau$ {\it vs} $\phi$ for  $N=64$ (triangle) and
      $N=256$ (circle).}
    \label{Fk_med_N64}
 \end{center}       
\end{figure}

\begin{figure}
  \begin{center}
    \vspace{-0.5cm}
    \rotatebox{-90}{\resizebox{5.9cm}{!}{\includegraphics{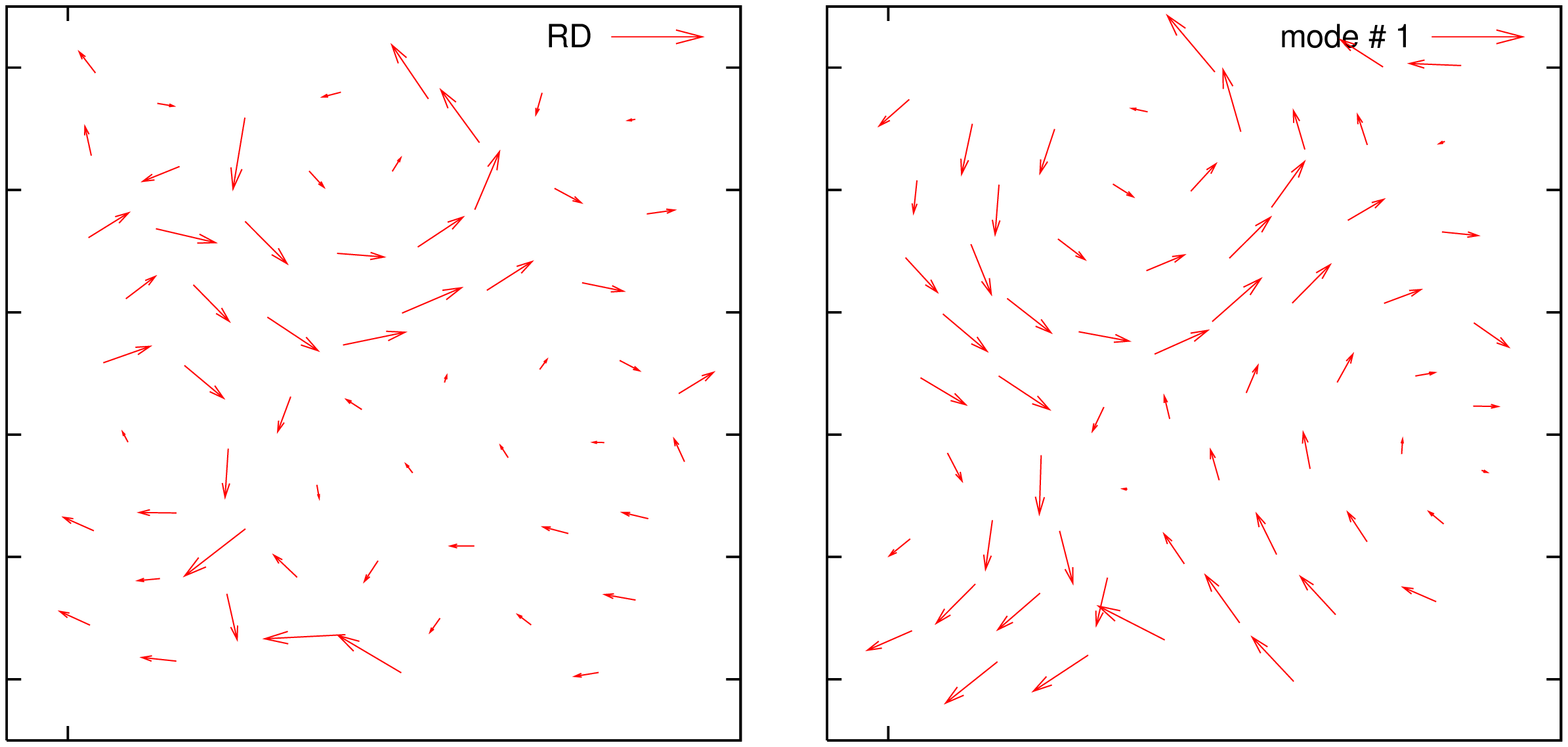}}}
    \rotatebox{-90}{\resizebox{5.9cm}{!}{\includegraphics{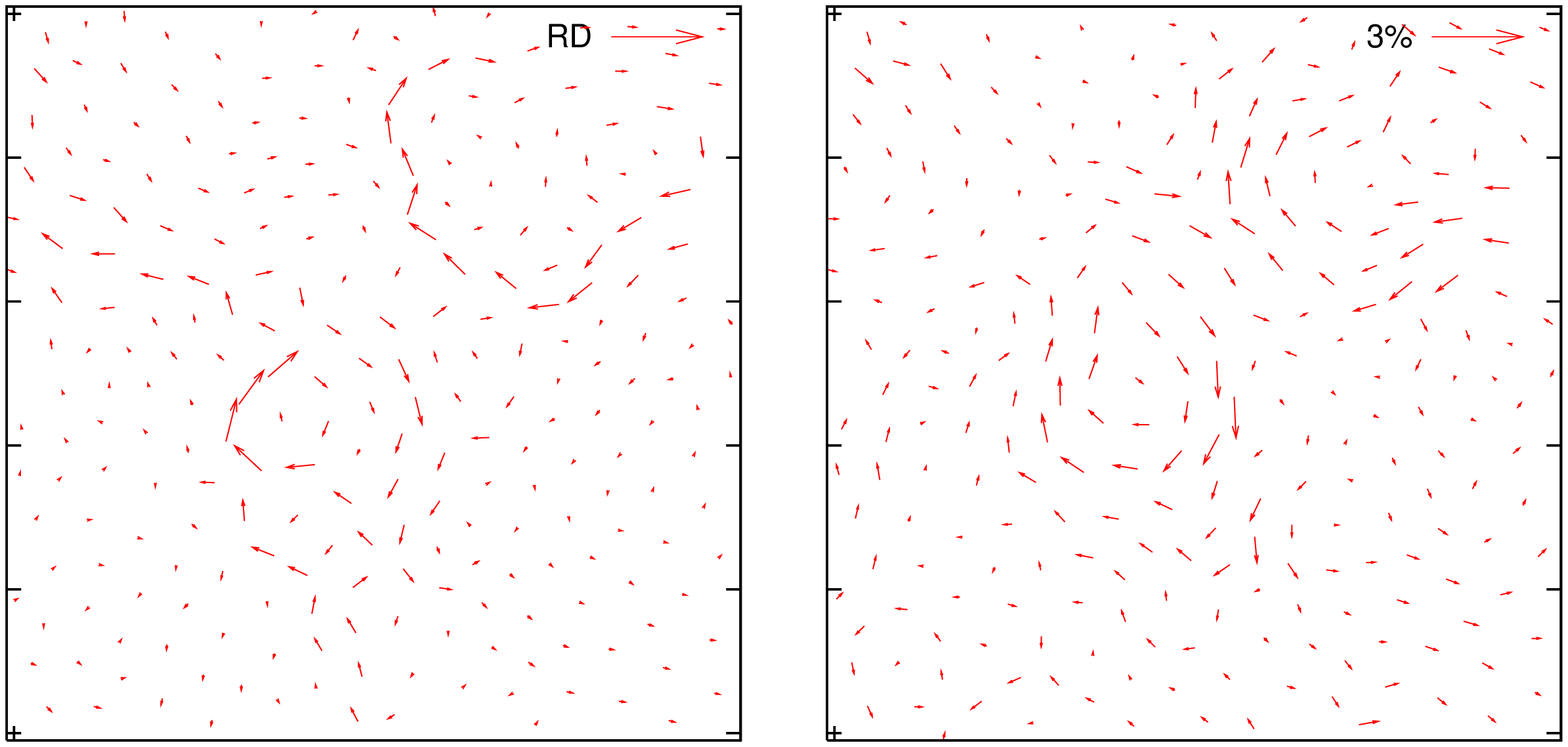}}}
    \vspace{-0.5cm}
    \caption{(a) Displacement field corresponding to the relaxation of
      Fig.(\ref{Fk_med_N64}-a). 
      Arrows were multiplied by 1.2.
      (b) lowest frequency normal mode,  which has the highest
      projection in this particular case. In this example one mode
      contributes to most of the displacement. (c) Relaxation
      event in system of $N=256$ particles at $\phi=0.786$. (d) Projection of the displacement (c) on the
      3\% of the mode that contribute the most to it.}
    \label{RD_AVEC_eps0.032}
 \end{center}       
\end{figure}

We now extend this analysis to an equilibrated super-cooled liquid. We
equilibrate at $0.77\leq\phi\leq0.786$. 
As previously observed e.g. in \cite{kob}, the dynamics  
is heterogeneous in time, and sudden 
rearrangements still occur on time scales of the 
order of $\tau$, the time scale of the $\alpha$-relaxation 
\footnote[2]{We define $\tau$ as the time for which $C(\vec q, \tau) = 0.3
  $.}. We use the procedure previously  described to
  determine  the average configuration of meta-stable
states,  and to define the displacement relaxing the structure,  
see Fig.(\ref{Fk_med_N64}-a).  We start by studying five packing fractions in a system of $N=64$ particles. 
For this size rearrangements generally span the entire system. A similar observation was made in a 3-dimensional lennard-Jones system of 125 particles \cite{kob}.   For each packing fraction, $F(k)$ and $F_{1/2}$ are computed for  seven
relaxation events.  
Results for $\langle F_{1/2}\rangle$ are presented in
  Fig.(\ref{Fk_med_N64}-b)   as a function of packing fraction. 
We find that  $\langle  F_{1/2}\rangle\leq 4\%$ for all $\phi$ studied,
implying that also in this region of the super-cooled liquid phase a small fraction
of the low-frequency modes contribute to the relaxation events. 
Interestingly, this fraction decays  significantly as the relaxation time grows,
suggesting a rarefaction of the number of directions  along which the system
can  yield near the glass transition. For the  largest $\phi\approx0.786$ we studied, $\langle  F_{1/2}\rangle\leq 2\%$, which implies that the collective event relaxing the system corresponds mainly to one or two modes. Fig.(\ref{RD_AVEC_eps0.032}-a) and Fig.(\ref{RD_AVEC_eps0.032}-b) compare one event and the mode that contributes most to it, which turns out to be  the softest mode in this particular example.

{\it Size effect}:   We now consider a  system of $N=256$ particles. An example of relaxation is shown in  Fig.(\ref{RD_AVEC_eps0.032}-c).  We observe for this system size that  a  larger fraction of the particles stay motionless. It is also  interesting to compare  for the same system size the aging dynamics deeper in the glass phase, e.g.   Fig.(\ref{ex_plat_crack}-b) with Fig.(\ref{RD_AVEC_eps0.032}-c): as was previously observed in LJ systems \cite{kb}, the collective rearrangements at equilibrium  involve less particles than earthquakes, but move them more.  Nevertheless in the equilibrated case as well, we shall see now that the observed displacements projects 
on a very small fraction of the normal modes. To study this question we perform the analysis introduced above, and compute $\langle F_{1/2}\rangle$ by averaging on  twelve events for the five packing fractions considered. Results are shown in Fig.(\ref{Fk_med_N64}-b): they are qualitatively similar to the system with $N=64$, but the values of $\langle F_{1/2}\rangle$ are  larger by approximately 0.5\% .  Nevertheless, it is well know that the glass transition occurs at lower packing fraction for smaller systems \cite{yy}, see inset of Fig.(\ref{Fk_med_N64}-c). When plotted versus relaxation time, the difference between the values of  $\langle F_{1/2}\rangle$ in the two systems is reduced and less systematic,  and $\langle F_{1/2}\rangle$ becomes in fact  smaller for $N=256$. Thus, even when embedded in a  system containing quiet regions, relaxation occurs along the  softest modes.

In kinetically constrained models \cite{andersen}, that have been proposed as paradigms for glassy dynamics, heterogeneous dynamics can arise from 
simple local microscopic rules. Our result that collective rearrangements correspond mostly to a few (and therefore necessarily reasonably extended) modes supports an
alternative view: elementary relaxations events are already extended objects, as are the soft  degrees of freedom of the system. 
We now argue that these modes are the anomalous modes \cite{matthieu1,matthieu2} described in the introduction.
Our justification for this lies in the microscopic structure of the glass: as evoked above, imposing the marginal stability of these modes leads to a power-law relation between coordination and pressure that is indeed observed in the glass phase \cite{caca}. If other soft objects (e.g. local configurations particularly soft due to disorder) were driving the dynamics, it is unlikely that the glass would freeze in this specific region of the phase space \footnote[3]{This region may correspond to the critical surface of mode coupling theory,
whose definition also depends on coordination, see \cite{lapo,kb} and reference therein.}.  

More needs to be known on the statistical properties of the anomalous modes, for example on the curly aspect of their displacement field or on their apparent capacity to form intense flow lines or strings. This problem turns out to be equivalent  to the statistics of force chains in amorphous solids \cite{tom}, a much studied problem in the granular literature,
but which still lacks a definite answer. Nevertheless, we know that
the anomalous modes are characterized by some length scale $l^*\sim p^{1/2}$ above
which the softest modes must extend \cite{matthieu1}.   The
observation that the softest modes dominate the relaxation supports
that regions of size at least $l^*\sim p^{1/2}\sim(\phi_c-\phi)^{-1/2}$ must rearrange. Very recent numerical work of sheared system near them jamming threshold  \cite{cond} supports our views, as in this case rearrangements are characterized by a diverging length scale with an exponent 0.6+/- 0.1, in agreement with our predictions.
These are also
consistent with the growing dynamical length scale observed near the
glass transition. Nevertheless  in this case the length scales at play,  typically about 5 or 10 particles, are too small to compare different theories \cite{durian}.
More stringent tests could be performed near maximum packing, where a diverging length scale is expected. This  could  be tested experimentally, e.g. by considering the intermittent aging dynamics of colloids  at large osmotic pressures.

Our work supports a unified description of both the structural relaxation and the packing geometry, where the dynamics corresponds to the collapse of anomalous modes,
and where the microscopic structure is fixed by their marginal stability. Note that the theoretical framework used here applies identically in three dimensions, where we expect our results to be valid as well. This scenario may also hold in other glasses, for example in Lennard-Jones where anomalous modes are also present \cite{ning}.
Nevertheless in this case, as for any long-range interaction potentials, $l^*$ is bounded above and does not diverge in the glass phase. 

We thank J-P. Bouchaud, L. G. Brunnet,  D. Fisher,  O. Hallatschek, S. Nagel
and T. Witten for helpful discussion and L.Silbert for furnishing the initial 
jammed configurations. C. Brito was supported by CNPq and M. Wyart by the Harvard Carrier Fellowship.


\begin{thebibliography}{99}

\bibitem{caca} C. Brito and M. Wyart, Europhys. Lett., {\bf 76} (1), pp. 149-155 (2006) 

\bibitem{matthieu1}  M. Wyart, S.R. Nagel, T.A. Witten, Europhys. Lett., {\bf 72}, 486-492, (2005)

\bibitem{matthieu2}  M. Wyart, L.E.Silbert, S.R. Nagel, T.A. Witten,  Phys. Rev. E {\bf 72} 051306 (2005)

\bibitem{cond} P. Olsson and S. teitel, cond-mat 07041806

\bibitem{simu} see e.g. M.M. Hurley and P. Harrowell, Phys. Rev. E 52, 1694 (1995); Y. Hiwatari and T. Muranaka, J. Non-Cryst. Solids, {\bf 235-237}, 19 (1998);  A. Widmer-Cooper and P. Harrowell, Phys. Rev. Lett. {\bf 96}, 185701 (2006) 

\bibitem{exp} M. T. Cicerone and M. D. Ediger, J. Chem. Phys. {\bf 103}, 5684 (1995); L. Berthier, G. Biroli, J-P. Bouchaud, L. Cipelletti, D. El Masri, D. L'Hote, F. Ladieu, M. Pierno; Science {\bf 310}, 1797 (2005).

\bibitem{weitz}  E.R. Weeks, J.C. Crocker, A.C. Levitt, A. Schofield, D.A. Weitz, Science, {\bf 287} (5453): 627-631 (2000)

\bibitem{glotzer} C. Donati, J. F. Douglas, W. Kob, S. J. Plimpton, P. H. Poole, and S. C. Glotzer, Phys. Rev. Lett. {\bf 80}, 2338 (1998);  S.C. Glotzer,  J. Non-Cryst. Solids, {\bf 274} (1-3): (2000) 

\bibitem{kob} G. A. Appignanesi, J. A. Rodriguez Fris, R. A. Montani, and W. Kob, Phys. Rev. Lett. {\bf 96}, 057801 (2006) 

\bibitem{toni} C. Toninelli, M. Wyart, L. Berthier, G. Biroli, and J-P. Bouchaud, Phys. Rev. E {\bf 71}, 041505 (2005)

\bibitem{tarjus} G. Tarjus, S.A. Kivelson, Z. Nussinov, and P. Viot, J.  Phys.-Cond. Matter {\bf 17} (50): R1143-R1182  (2005) 

\bibitem{phillips_book} {\it Amorphous Solids, Low Temperature Properties}, edited by W. A. Phillips (Springer, Berlin, 1981).

\bibitem{alexander} S.Alexander,  Phys. Rep.,{\bf  296}, 65 (1998)

 \bibitem{parisi} G. Parisi, Eur. Phys. J.E., {\bf 9} 213-218 (2002)

\bibitem{keyes} T. Keyes, {\it J. Phys. Chem. A} {\bf 101}, 2921 (1997).

\bibitem{saddle} T.S. Grigera, A. Cavagna, I. Giardina, G.Parisi, Phys. Rev. Lett., {\bf 88}, 055502 (2002); D. Coslovich and G. Pastore, Europhys. Lett., {\bf 75} (5), 784-790 (2006)

\bibitem{chandler} D. Chandler, J.D. Weeks, H.C. Andersen, Science {\bf 220} (4599): 787-794 1983

\bibitem{matthieu3} M. Wyart, {\it Ann. Phys. (Paris)}, {\bf 30}, No. 3 (2005) pp.1-96, or arXiv cond-mat/0512155

\bibitem{J} C.S O'Hern, L.E Silbert, A. J. Liu and S.R. Nagel,  Phys. Rev. E,  {\bf  68}, 011306 (2003)
  
\bibitem{sil} L. E. Silbert, A. J. Liu, and S. R. Nagel, {\it Phys. Rev. Lett} {\bf 95}, 098301 (2005).

\bibitem{dove} K. Trachenko, M.T. Dove, V. Brazhkin and F.S. El'kin  Phys. Rev. Lett. {\bf  93}, 135502 (2004)

\bibitem{ning}  N. Xu, M. Wyart, A. J. Liu, S. R. Nagel, arxiv cond-mat 0611474


\bibitem{bubble} A. Ferguson, B. Fisher, B. Chakraborty, Europhys. Lett., {\bf 66}, 277 (2004)

\bibitem{kb} W. Kob and J-L. Barrat, Eur. Phys. J. B {\bf 13}, 319-333 (2000)

\bibitem{lapo}  W. Kob W, JL. Barrat, F. Sciortino,. P. Tartaglia J., Phys. Condensed Matter {\bf 12} 6385 (2000)

\bibitem{duri} A.Duri, P Ballesta, L. Cipelletti, H. Bissig and V. Trappe, Fluctuation and Noise Lett.,{\bf 5}, 1-15, (2005); L Buisson, L Bellon and S Ciliberto, J. Phys.: Condens. Matter {\bf 15} S1163ÐS1179  (2003)

\bibitem{as} Neil Ashcroft and N.David Mermin, {\it Solid
  state physics}, New York (1976).

\bibitem{caca2} C. Brito and M. Wyart, in preparation

\bibitem{yy} K. Kim and R. Yamamoto, Phys. rev. E, {\bf 61}, R41, (2000)

\bibitem{andersen} G.H. Fredrickson and H.C. Andersen, Phys. rev. Lett., {\bf 53}, 1244 (1984); W. Kob and H.C. Andersen, Phys. Rev. E {\bf 48}, 4364 (1993).

\bibitem{tom} A.V. Trachenko and T.A. Witten, Phys. Rev. E {\bf 60}, 687 (1999); A.V. Trachenko and T.A. Witten, Phys. Rev. E {\bf 62}, 2510 (2000)

\bibitem{durian} see e.g. A. S. Keys, A. Abate, S. C. Glotzer, and D. Durian, Nature Physics, doi:10.1038/nphys572



\end{thebibliography}
\end{document}